\documentclass[preprint]{aastex}
\usepackage{graphicx}

\shorttitle{Radio Bursts from the Crab Pulsar}
\shortauthors{Crossley et al.\ }

\begin{document}
\newcommand{\mus}{\,$\mu$s}
\newcommand{\mctc}{\multicolumn{2}{c}}
\def\ltw{\>\hbox{\lower.25em\hbox{$\buildrel<\over\sim$}}\>}
\def\gtw{\>\hbox{\lower.25em\hbox{$\buildrel>\over\sim$}}\>}

\title{Short-lived Radio Bursts from the Crab Pulsar}
\author{J. H. Crossley\altaffilmark{1}, J. A. Eilek\altaffilmark{2,3},
 T. H. Hankins\altaffilmark{2,3}, J. S. Kern\altaffilmark{1}}
\altaffiltext{1}{National Radio Astronomy Observatory, Socorro, NM 87801}
\altaffiltext{2}{Physics Department, New Mexico Tech, Socorro, NM 87801}
\altaffiltext{3}{Adjunct Astronomer, National Radio Astronomy Observatory}
\email{jeilek@aoc.nrao.edu}

\begin{abstract}
  Our high-time-resolution observations reveal that individual main
  pulses from the Crab pulsar contain one or more short-lived
  microbursts. Both the energy and duration of bursts measured above 1
  GHz can vary dramatically  in less than a millisecond.  These
  fluctuations are too rapid to be caused by propagation through
  turbulence in the Crab Nebula or the interstellar medium; they must
  be intrinsic to the radio emission process in the pulsar.  The mean
  duration of a burst varies with frequency as $\nu^{-2}$,
  significantly different from the broadening caused by interstellar
  scattering. We compare the properties of the bursts to
  some simple models of microstructure in the radio emission region.

\end{abstract}


\section{Introduction} 

As part of the quest to understand how pulsars make coherent radio
emission, we must understand how available energy in the pulsar's
magnetosphere is converted to radio emission.  For instance,
unshielded electric fields can drive relative streaming of electrons
and positrons, either in outflows from the polar caps or in ``gap''
regions in the upper magnetosphere. The energy of these plasma streams
is, in principle, available to be converted to radiation.  Energy
stored in disordered magnetic fields may also be available to be
converted to radio emission.  The relevant physical scales are
probably large (tens of km or more, corresponding to millisecond
timescales), and are manifested by pulsar mean profiles which reveal
the radio-loud regions in the magnetosphere.  On very small scales,
this available energy is somehow converted to the collective charge
motions which actually create the coherent radio emission.  We expect
this to take place on scales comparable to the plasma scale (typically
tens of cm).  An example of this may be the ``nanoshots'', 
  lasting no longer than a nanosecond, which Hankins et al.\ (2003)
found in the main pulse of the Crab pulsar (also see Hankins \& Eilek
2007).  These nanoshots may reveal the fundamental emission process in
their region of origin;  one likely model is soliton collapse in strong plasma
turbulence (Weatherall 1998).  

However, we do not yet understand how these two spatial scales are
coupled.  Just how are plasma dynamics on magnetospheric scales 
  (hundreds of km) connected to the very small scales  (tens of
  cm)  involved in coherent radio emission?  We suspect the
answer will come from {\it microstructure}, which we define as
significant pulse structure on $\sim 1$ to 100 $\mu$s timescales
(corresponding to $0.3$ to 30 km spatial scales).

\subsection{Models of microstruture}
\label{sec:intro:micro}

Microstructure has been known to exist almost since pulsars were
discovered (Craft et al.\ 1968, Hankins 1971). 
Although a variety of models have been proposed to explain the phenomenon, 
none has  emerged as definitive.

Early models tended to be based on geometry.  Various authors have
proposed narrow, long-lived structures (flux tubes, say) which exist
within the open field line region.  These structures radiate by
coherent curvature radiation as they rotate past the line of sight
(e.g., Benford 1977; Lange et al.\ 1998).  In this picture, the
microsecond timescales reflect the size or beaming angle of the
radio-loud structures.  A variant of this model proposes
 localized charge clouds
which intersect the line of sight only briefly as they move out
along open field lines (e.g., Cordes 1981; Gil \& Melikidze 2005).
The timescales then reflect the time during which a given charge cloud
beams its radiation into the observer's line of sight.

Alternative models have also been proposed.  Microstructure has been
suggested to come from intrinsic temporal variability of the radio
emission, possibly due to dynamic instabilities or unsteady flow in
the emitting plasma (e.g., Sheckard, Eilek \& Hankins 2010), for
instance by magnetically driven flares (Lyutikov 2003).  Timescales
here would be those of the underlying instability (such as magnetic
reconnection).  Still another class of model identifies microstructure
with propagation effects in the magnetosphere.  Petrova (2004)
proposes that microstructure is caused by the variable gain in
stimulated Compton scattering; the timescale reflects the angular
width of the high-gain radiation beam as it rotates past the line of
sight.  Cairns, Johnston \& Das (2003) suggest microstructure comes
from the stochastic growth of a signal passing through a turbulent
plasma.

\subsection{Goal of this work}

To discriminate among these theories, we need observations that can
confront the models. Traditionally microstructure has been treated
statistically (e.g., Hankins 1972; Cordes, Weisberg \& Hankins 1990;
Lange et al.\ 1998), but such methods can obscure important details of
the process.  We therefore turn to sensitive 
observations of individual pulses at sub-microsecond time resolution.

The Crab pulsar is an opportune target for single-pulse microstructure
studies.  Its occasional very bright pulses (by which it was
discovered; Staelin \& Reifenstein 1968) are good targets for our
high-time-resolution data acquisition system (Hankins et al.\ 2003).
At frequencies above $\sim 1$ GHz, single pulses from the Crab pulsar
consist of one to several  components, each lasting $ \sim 1\! -\! 100$
 microseconds. We call these {\it microbursts} (or simply
{\it bursts}).  In many pulses these bursts are sufficiently far apart
in time that we can identify and characterize each one.  Sallmen et
al.\ (1999) show a few examples of single pulses at 1.4 GHz, taken
from two of the data sets we analyze in this paper; Hankins \& Eilek
(2007) show a few examples observed at 9 GHz. As we show here, the
properties of these bursts are highly variable.  The number of bursts
varies from one pulse to the next; the rotational phase, energy, and
duration of a burst can vary between consecutive pulses and even
within a single pulse.

In this paper we describe the microbursts we have detected in
high-time-resolution, single-pulse observations of the main pulse of
the Crab pulsar.  We restrict ourselves to the main pulse for two
reasons.  First, pulses bright enough to be detected by our data acquisition
system at frequencies below 5 GHz are much more common at the
rotational phase of the main pulse than that of the interpulse (Cordes
et al.\ 2004).  We  recorded some bright interpulses in a few VLA
observing sessions at 1.4 and 4.8 GHz.  Inspection of these data shows
that these interpulses also contain microsecond-long bursts, very
similar to those of the main pulse which we present in this paper.
However, bright interpulses below 5 GHz are too rare in our data to
constitute a statistically significant sample. In addition, the
interpulse of the Crab pulsar seems to change its character, and its
rotational phase, at high radio frequencies. The phase of the radio
interpulse remains steady from 0.1 to 1.4 GHz, and is consistent
with phase of the interpulse at optical and X-ray frequencies (Rankin
et al.\ 1970; Moffett \& Hankins 1996);  but the interpulse
seen between $\sim 5$ and 10 GHz occurs at a slightly earlier
rotational phase (Moffett \& Hankins 1996). Both the
temporal and spectral characteristics of the interpulse between 
$\sim 5$ and 10 GHz are dramatically different from those of the main pulse
at the same frequencies (Hankins \& Eilek 2007).  We therefore
restrict ourselves here to the main pulse of the Crab pulsar.

Our high-time-resolution observations need the brightest pulses, which
have been loosely called ``giant'' pulses in the literature.  There
has been discussion as to whether or not such bright pulses from the Crab
pulsar are typical of the general pulse population (and thus, whether
or not conclusions drawn on the basis of bright pulses can be applied
to general pulsar emission physics).  Lundgren et al.\ (1995)
suggested that giant pulses at 0.6 GHz are a separate population from
more typical, fainter pulses.  Neither Popov \& Stappers (2007) nor
Sheckard et al.\ (2010) found any sign of a bimodal pulse distribution
at 1.4 GHz,  but Karuppusamy et al.\ (2010) find a possible
  second population at their fainter sensitivity limits.  We do not
pursue this question in this paper, but note that the pulses we study
here correspond to the high-energy tail of bright-pulse distributions
measured by these authors.

In the rest of this paper we present our data and characterize the
microburst distribution of the main pulse of the Crab pulsar. In
\S\,\ref{sec:obs} we describe the observations; in \S\,\ref{sec:fit}
we describe the functional fitting we use to measure the bursts.  In
\S\,\ref{sec:energies} we discuss the variabilty and energetics of the
bursts. In \S\,\ref{sec:widths} we discuss the durations of the
bursts, and show that above $\sim 1$ GHz the burst widths are
intrinsic to the star.   We conclude with a short discussion  
in \S\,\ref{sec:conclusions}.

\section{Observations} 
\label{sec:obs}

We recorded single pulses from the Crab pulsar in several observing sessions
at the Very Large Array (VLA)\footnote{The
  Very Large Array is an instrument of the National Radio Astronomy
  Observatory, a facility of the National Science Foundation operated
  under cooperative agreement by Associated Universities, Inc.} 
  between 1993 (MJD 49080) and 1999 (MJD 51218).  We used the VLA    
in  phased-array mode, in which 
individual antenna delays are set in real-time so as to restore the
original wave front.  Even in the short-baseline, D-array configuration, 
the VLA at
both 1.4 and 4.9\,GHz synthesizes such a small angular beam that
it is insensitive to most of the structure of larger
angular size in the Crab Nebula when pointed at the pulsar.
We therefore attained a much lower system temperature on
the Crab pulsar than is possible with a single-dish telescope of
equivalent collecting area.

The software package TEMPO (Taylor \& Weisberg 1989) was used in
  prediction mode 
to create a pulsar timing model based on the monthly Crab pulsar
ephemeris published by Jodrell Bank 
Observatory\footnote{\url{http://www.jb.man.ac.uk/$\sim$pulsar/crab.html}}  
(Lyne, Pritchard \& Graham-Smith   1993).
The timing model was  used to set a period-synchronous gate at the pulse
phase of the Crab main pulse.  For each pulsar
period the signal within the gate was square-law detected with a
200-$\mu$s time constant.  If the average flux of the  pulse
within this gate exceeded a preset multiple (typically 7) of the
root-mean-square off-pulse noise, the pulse was recorded:  the  two
orthogonal, circularly polarized signal voltages were digitally
sampled by a LeCroy oscilloscope at a 
rate of 100 MHz.  The samples were then transferred to disk for
subsequent off-line coherent dedispersion (Hankins 1971).  During
the data transfer time, 10--30\,s, the data acquisition was disabled, and
no pulses could be captured. 

In this paper we consider only those observing sessions in which no
fewer than 27 strong main pulses were recorded.  This number was chosen, after
examining the data, to include a useful and interesting ensemble of
observing sessions but to avoid observing sessions with statistically
insignificant numbers of pulses.

This leaves us with 26 VLA observing sessions, listed in 
Table \ref{tab:VLA obs table}.  Most of our observations were made 
between 1.2 and 1.7 GHz, or between 4.5 and 5.0 GHz;  we include
one observing session at 0.33 GHz.  All observations were made
with 50-MHz bandwidth, with two exceptions:  the lowest frequency
observations allowed only 3.125-MHz bandwidth at 0.33-GHz center
frequency,  and a simultaneous two-frequency observation at 1.2 and
1.7 GHz used 25-MHz bandwidth at each frequency.  Table \ref{tab:VLA obs table}
 also lists the number of pulses and microbursts fitted (as described in 
\S\,\ref{sec:fit}), the smoothing time used for the fits, the span of
time between first and last recorded pulse, and the MJD for each
observing session.  In total, the VLA observations reported here include
1551 Crab pulses which contain 2969 fitted microbursts.

Seven of the observing sessions listed in Table \ref{tab:VLA obs table}
were simultaneous two-frequency observations, in which the VLA was
split into two independent sub-arrays. Since interstellar dispersion
causes the lower frequency signal to arrive later, we set the
period-synchronous gate to monitor the pulse energy of the
higher-frequency sub-array.  If a pulse was recorded there, the
oscilloscope was triggered again, after the appropriate digitally
controlled dispersion delay, to record the pulse again in the lower
frequency band.  These seven pairs of observing sessions can be
identified by the superscripts on the first column of Table \ref{tab:VLA
  obs table}; corresponding superscripts denote simultaneously
recorded data.

We also used the same data acquisition system and observing procedure
to record single pulses from the Crab pulsar in a few observing
sessions at Arecibo Observatory\footnote{Arecibo Observatory is
  part of the National Astronomy and Ionosphere Center, which is
  operated by Cornell University under a cooperative agreement with
  the National Science Foundation.} in 2002.  These are listed in
Table \ref{tab:AO obs table}.  We did not carry out our full microburst
analysis on these data sets, but we do use them in our statistical
analysis of microburst widths, discussed in \S \ref{sec:widths}.

\section{Microburst Function Fitting}
\label{sec:fit}

Figure \ref{fig:fit example} shows several examples of individual pulses
we recorded between 0.33 and 4.9 GHz.  These examples are
characteristic of the full data set.  At all frequencies above 0.33 GHz 
single pulses contain one to several microbursts which are
sufficiently well separated to be identified individually.  The bursts
can vary significantly in amplitude, duration, arrival phase, and
total energy, from pulse to pulse and even within a given pulse.

\subsection{Fitting function and procedure}
\label{sec:fits}

To analyze the microbursts, we fitted analytic functions to individual
bursts in each pulse.  Inspection of our data showed that a
  fast-rise, slow-decay function, $F(t) \propto t e^{-t / \tau}$ (as
  illustrated in Figure \ref{fig:fit example}), is a good match to
  most of the individal bursts (as we discuss in \S
  \ref{sec:fit_robust}).  We therefore chose this as our fitting
  function. Our fitting procedure was as follows.

\begin{figure}[]
{\center
\includegraphics[width=0.6\textwidth]{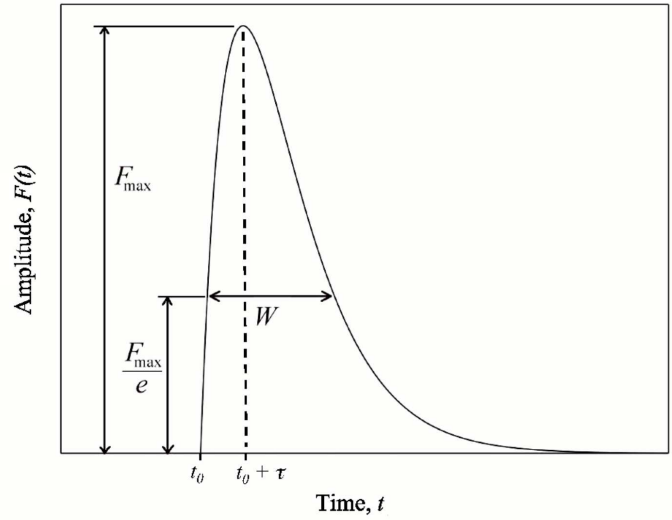}
\caption[]{Illustration of the microburst fitting
  function, $F(t) = A (t-t_0) e^{-(t-t_0)/\tau}$.  The function reaches a
maximum,  $F_{\rm max} = A \tau e^{-1}$, at a time $t = t_0 + \tau$.
  Its width, defined as $W= 3 \tau$, is approximately
 the width at an amplitude of $F_{\rm max}/e$.
  The total fluence is ${\cal E}  = \int F(t) dt = A \tau^2$.}
\label{fig:xfred example} }
\end{figure}

After coherent dedispersion (Hankins 1971), the off-pulse mean
intensity, $F_{\rm off}$, 
 is calculated and subtracted from each individual pulse.
We note that $F_{\rm off}$  consists of
 the receiver system temperature and a  contribution  from the Crab Nebula.
We fit the remaining pulse flux as a sum of  $N$ component bursts,
\begin{equation}
F(t) = \sum_{i=1}^N F_i(t; A_i, \tau_i, t_{0i}) 
\label{xfred_sum}
\end{equation}
where the $i^{\rm th}$ burst is
\begin{equation}
F_i(t; A_i, \tau_i, t_{0i}) = A_i (t - t_{0i})\, e^{-(t -   t_{0i}) / \tau_i}.
\label{xfred}
\end{equation}
This function is
illustrated in Figure \ref{fig:xfred example}, and examples of fitted
microbursts are shown by dashed lines in each pulse of Figure \ref{fig:fit
  example}.
Each burst, $F_i$, is described by three parameters: its start
time $t_{0i}$, its amplitude $A_i$, and its decay time $\tau_i$.  From
these, we derive the burst's maximum flux $F_{{\rm max}, i} = A_i
\tau_i e^{-1}$, which occurs at its time-of-arrival, $t_{{\rm TOA}\,i}
= t_{0i}+\tau_i$.  In what follows we measure the 
burst's duration by its full width, $W_i = 3
\tau_i$, which is approximately equal to the length of time that the flux
exceeds $F_{{\rm max}, i}/e $.  In the rest of this paper, we drop the ``max''
subscript and use ``$F$'' to denote the maximum flux of a burst.

The total energy of a burst is related to the time-integrated flux, 
which we call the {\it fluence}: $ {\cal E}_i =
  \int F_i(t) dt = A_i \tau_i^2 = F_i W_i e / 3$.  To
  convert ${\cal E}$ to total burst 
energy, $E$, we must know the distance to the
  pulsar, $D$; the bandwidth of the radiation, $\Delta \nu$;
  and the angular width of the radiation beam, $\theta$: $E = {\cal E}
  \Delta \nu D^2 \theta^2$.  We present most of our results here in
  terms of the fluence, ${\cal E}$, which is the directly measured
  quantity;  we discuss the underlying energy release in \S
\ref{subsec:energies} and \S \ref{subsec:widths}, below.

We used up to six terms in our fits,
 ($1 \le N \le 6$), as needed for each pulse, so that
uncertainties in the function parameters were small and the
pulse residuals were close to being normally distributed about zero
with standard deviation $\sigma(t)$, as given by the radiometer equation. 
For the on-pulse region, that standard deviation is
\begin{equation}
\sigma_{\rm  on} (t) =  
{[F(t) + F_{\rm off}] /\sqrt{\Delta\nu\,\Delta t} }
\end{equation}
while for the off-pulse region, 
\begin{equation}
\sigma_{\rm  off} =  
 F_{\rm off} / \sqrt{\Delta\nu\,\Delta t} 
\end{equation}
where $\Delta\nu$ is the observing bandwidth and  $\Delta t$ is the
smoothed time resolution.

For  each   pulse,  the  microbursts  were   initially  identified  by
inspection, and the three parameters, $A_i, t_{0i},$ and $\tau_i$, were 
carefully 
estimated by eye. We then  used $\chi^2$ minimization to  obtain the
best fit of $F(t)$ to the data.  In computing $\chi^2$ for a given set
of  parameters  $(A_i,  \tau_i, t_{0i},  i=1,N)$, every  $j$'th data
sample  must  be  assigned  a  standard  deviation,  $\sigma_{j}$,  to
represent  its uncertainty.   After  experimenting with  a variety  of
methods for calculating $\sigma_{j}$, we found that using the constant
off-pulse standard  deviation for all samples gives  fits with  the best
residuals   after   $\chi^2$   minimization.   Using   $\sigma_{j}   =
\sigma_{\rm  on}(t)$ as  the standard deviation {\it unweights}
  the strongest parts
of the  pulse and $\chi^2$  minimization yields poor fits.   Thus, the
standard  deviation  $\sigma_{j}$  used  to compute  $\chi^2$  is  
formally  the same  as  the standard  deviation, $\sigma_{\rm off}$;  after
$\chi^2$ minimization  we judge the fit according  to $\sigma_{\rm on}(t)$ to
avoid non-physical $\chi^2$ minima.

Although the fits were carried out using the full time resolution of
the data 
(usually 10 ns), we smoothed and
decimated the data before identifying the bursts, typically to 0.8
$\mu$s for pulses below 2 GHz, and to 0.2 $\mu$s at higher
frequencies.  Table \ref{tab:VLA obs table} gives the smoothing time used
for each data set.

\subsection{How robust are the microburst fits?}
\label{sec:fit_robust}

\begin{figure}[]
{\center
\includegraphics[width=.8\textwidth]{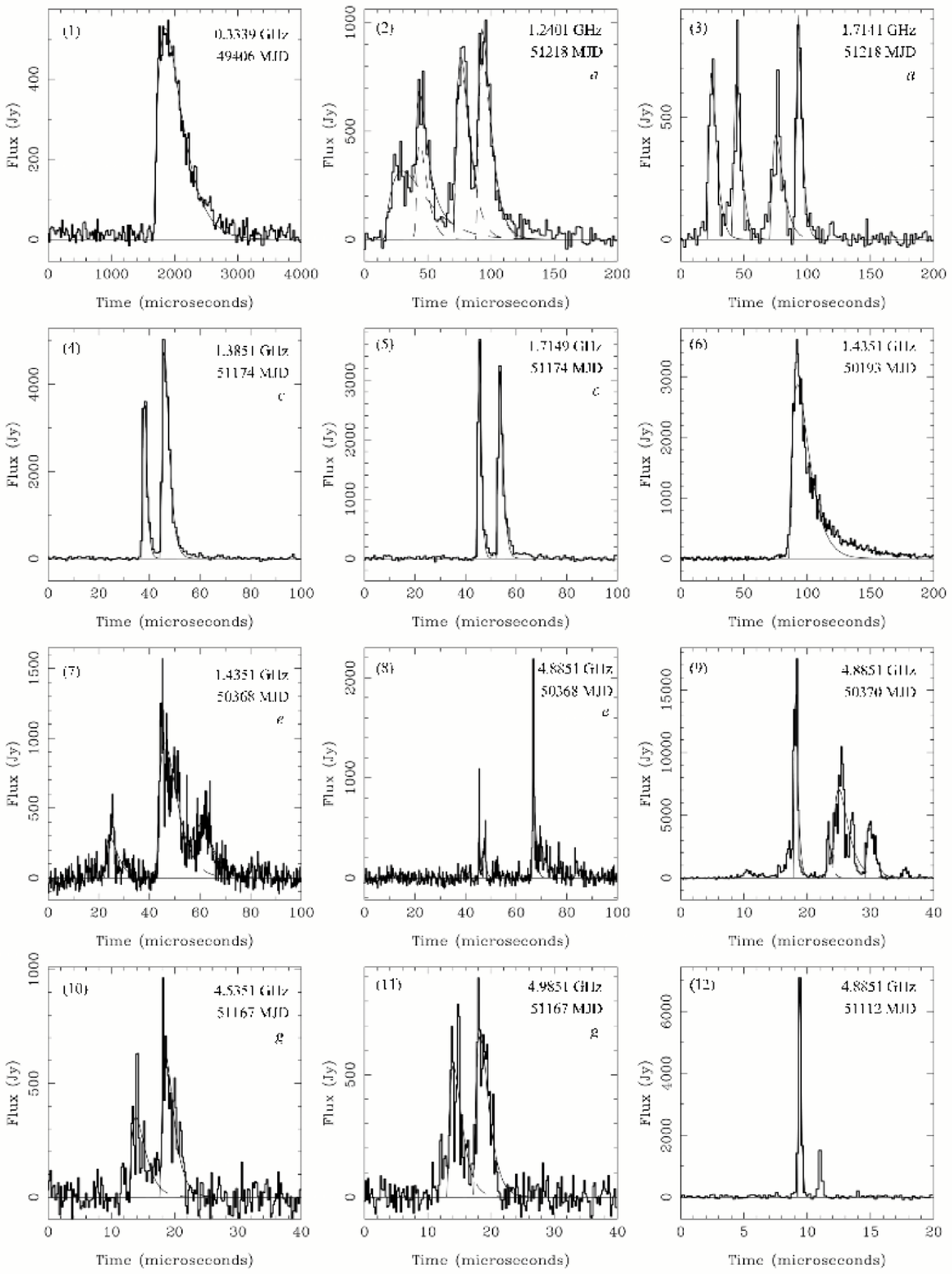}
\caption[]{Examples of single pulses recorded from the
  Crab pulsar.  These examples illustrate the fact that
several microbursts are often seen in an individual pulse, except at
0.33 GHz where interstellar scattering dominates the profile.  These
examples also illustrate  the gradual change of
  burst profiles from lower to higher frequencies.  Each pulse is
  identified by center frequency and observing date (MJD), as well as a
  reference number at the top left.  Pulses recorded simultaneously at two
  frequencies are identified by letters given below the observing dates;
these letters correspond to superscripts in Table \ref{tab:VLA obs table}. 
Heavy lines show the data; light dashed lines are 
the functional fits to individual bursts, discussed in
  \S\,\ref{sec:fit}; the light solid lines are their sum. }
\label{fig:fit example} }
\end{figure}

We find that the fast-rise, slow-decay function given in equation
(\ref{xfred}) fits the microbursts well at low frequencies.  At our
lowest frequency, 0.33 GHz, the single pulse profiles are very well
fitted with a single $F(t)$ function (i.e., $N=1$ in equation
\ref{xfred_sum}), as illustrated by pulse 1 in Figure \ref{fig:fit example}.
 This agrees with previous results;   Rankin et al.\ (1970) showed that
 the function in equation (\ref{xfred}) is a good description of
the  pulsar's mean profiles between 0.074 and 0.43 GHz.

At higher frequencies, however, the microburst shape begins to deviate
from the fast-rise, slow-decay shape of our fitting function. Between
1.2 and 1.7 GHz, microbursts are still well matched to our
fast-rise, slow-decay function, but they show occasional deviations from the
functional form at the beginning or end of the microburst.  For
example, see pulse 6 in Figure \ref{fig:fit example}.  Above 4.5 GHz,
the bursts show less of the slow-decay tail, and tend to be more
symmetric about their peak than bursts at lower frequencies.  In
addition, higher-frequency bursts can contain very narrow bursts of
emission superimposed upon broader features which may resemble the
fast-rise, slow-decay shape of our fitting function. Examples can be
found in pulses 9 and 10 in Figure \ref{fig:fit example}.

Our fitting procedure clearly misses narrow and weak bursts.  The
narrowest burst we can identify is determined by our smoothing time,
which is at least twenty times
larger than our intrinsic time resolution.  Because our
fitting procedure is less than robust when bursts overlap in
time, we can miss very narrow bursts which are superimposed on
a broader burst of similar strength.  We also clearly miss weaker
bursts, for two reasons.  First, we recorded only the stronger pulses,
because we triggered our data acquisition system on the  
fluence of a pulse. In addition, weaker bursts in a multi-burst pulse might not
be picked out by our fitting procedure.

\subsection{Bursts seen simultaneously at two frequencies}
\label{sec:two freq}

We compared individual pulses in the data sets which were recorded
simultaneously at two different frequencies (denoted by superscripts
in Table \ref{tab:VLA obs table}).  We determined by inspection that the
same microburst can be identified at both the high and low frequency
in some, but not all, frequency pairs.  We can identify the same burst
at two different frequencies if those two bands are both between 1 and
2 GHz, or both between 4 and 5 GHz.  Examples of the same bursts which
can be identified in pulses observed simultaneously at two frequencies
are shown in pulse pairs 2 and 3, 4 and 5 or 10 and 11 of Figure
\ref{fig:fit example}.

However, we could not reliably identify individual bursts at both
frequencies in pulses observed simultaneously at 1.4 and 4.9 GHz, as
exemplified by pulses 7 and 8 in Figure \ref{fig:fit example}.  Thus,
our fractional microburst bandwidth, $\Delta \nu / \langle \nu
\rangle$, is at least a factor $\sim 1/3$, but does not exceed
unity. At first glance this seems to contradict the result reported by
Moffett (1997), that 90\% of bright pulses detected at 4.9 GHz were
also detected at 1.4 GHz.  Similarly, Sallmen et al.\ (1999) found
that 70\% of bright pulses detected at 1.4 GHz were also seen at 0.6
GHz.  We conclude that although a bright pulse can be broadband 
  (extending to an upper frequency at least 2-3 times the lower
  frequency), the microbursts which comprise it are relatively
narrow-band.  Whatever the physical process that releases energy into
a bright pulse, it does so by means of a set of microbursts, each
emitting in its own band of radio-frequencies.

\section{Microburst variability and energy}
\label{sec:energies}

Our basic measured quantities are the flux, $F$, and width, $W$, of an
individual microburst.  As example pulses in Figure \ref{fig:fit example}
illustrate, both the flux and width of a burst can vary dramatically,
on timescales as short as several microseconds, even within a single
pulse.

\begin{figure}[]
{\center
\includegraphics[width=0.7\textwidth,angle=90]{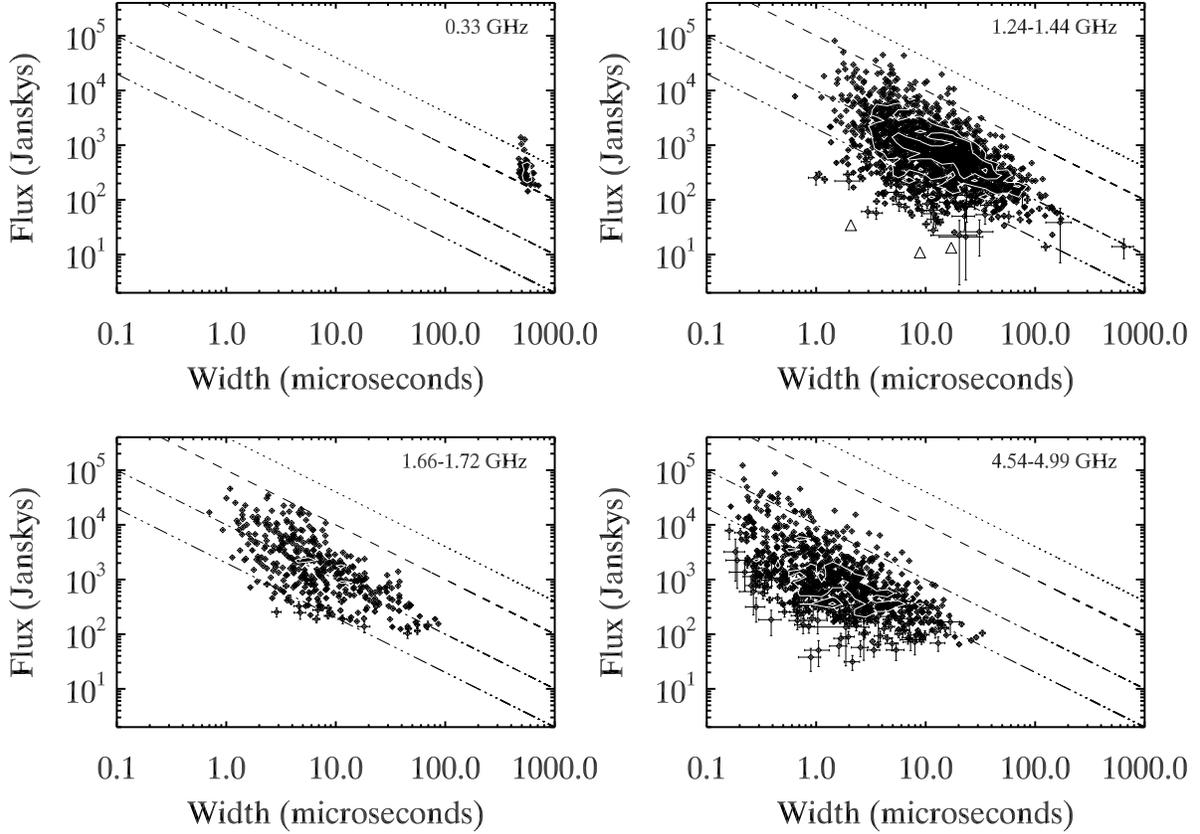}
\caption{Distribution of the measured flux and width values for
  all observed microbursts, grouped by frequency as labeled.
The overlaid lines
  show loci of constant  fluence: dotted, $0.4$ Jy-s; dashed, $0.1$
  Jy-s; dash-dot, $0.01$ Jy-s; triple-dot-dash, $0.002$ Jy-s. 
While bursts occupy a wide range of fluence, 
there is some tendency for brighter bursts to be short-lived,  and
 fainter bursts to be longer-lived.  The
contours show point densities of 5, 10 and 20 points per bin;  the bin size
is 0.1 decade in log space.} 
\label{fig:flux-width} }
\end{figure}

To show this, we first present our results as $(F,W)$ plots, grouped
by observing frequency, in Figure \ref{fig:flux-width}.  
Each point is plotted
with error bars denoting the fitting uncertainty in each quantity;
nearly all of the points have uncertainties no larger than the
plotting symbol used. (A very few data points, plotted as triangles,
do not have error bars because the uncertainties are large enough to
push the bounds of the error bars beyond the limits of one or both axes.)
Despite the  wide variation possible from burst to burst, Figure
\ref{fig:flux-width} shows that the distribution of
individual bursts is localized in $(F,W)$ space.  As discussed above,
in \S \ref{sec:fit_robust}, our method misses weak and narrow overlapping pulses;
 the lower and leftmost boundaries of the $(F,W)$ distributions
are not physical.  However, the lack of points in the upper right of
each plot is physical; we would easily have detected such strong, wide
bursts.

In this figure 
we have combined our data from different observing
epochs into frequency groups.  Although the flux and width can vary
widely from burst to burst, we determined by inspection of individual
data sets that the apparent ``centroid'' of the ($F,W)$ distribution,
at a given frequency, varies by no more than a factor of a few between
observing sessions.  Data from the individual epochs are presented in
Crossley (2009).  The small variation of mean fluxes we find agrees
with previous authors (Rickett \& Lyne, 1990, at 0.8 GHz, Lundgren et
al., 1995, at 0.61 GHz, and Rankin et al., 1974, 0.073 to 0.43 GHz)
who found the flux of time-averaged Crab pulses varied by only a
factor of a few on timescales of several days to a few hundred days.

\subsection{Fluence and energy range for bursts}
\label{sec:fluence_energy_range}

 It is well known that pulsar radio emission is highly variable
  from pulse to pulse, at least when observed at lower time
  resolution.  Our results show that this variability also holds at
  the microburst level.  

\begin{figure}[]
{\center
\includegraphics[width=0.7\textwidth,angle=90]{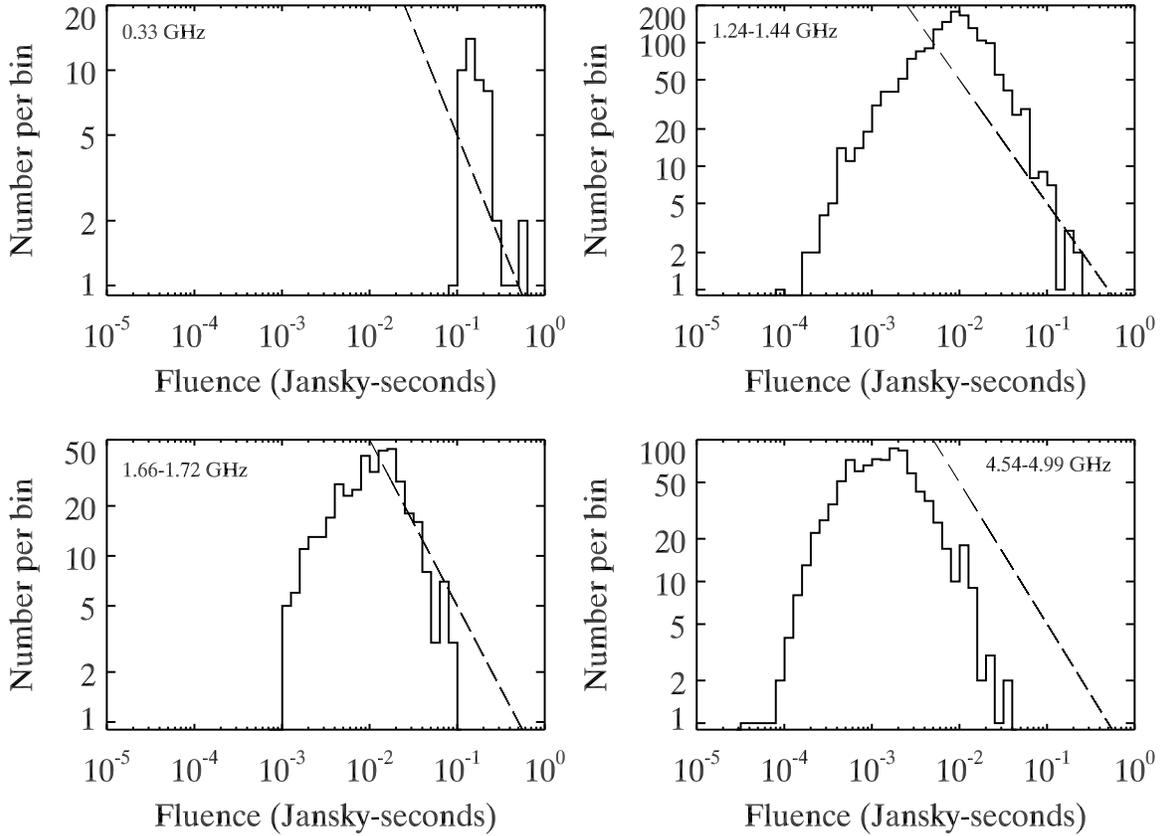}
\caption{ Distribution of fluences for all measured microbursts,
displayed as a histogram in log space, grouped in the same frequency ranges
used in Figure \ref{fig:flux-width}. The data in this figure are binned in
one-tenth-decade intervals; the histogram values are not normalized to
the bin size.  Thus, the units of the vertical axis are ``number of
microbursts'', not ``number of microbursts per fluence range''.   The 
apparent low-fluence falloff is a result of incomplete sampling 
due to the triggering mechanism we used to record single pulses.  The dashed
line illustrates the ${\cal E}^{-2}$ distribution which would result if all bursts
release the same amount of energy which is modified only by relativistic
beaming at different angles to the line of sight. }
\label{fig:energy hist} }
\end{figure}

  We overlay lines of constant fluence in each $(F,W)$ plot in Figure
  \ref{fig:flux-width}.  The fluence values of our bursts range over a
  factor $\sim 100$; we emphasize again that such variability can
  occur within a single pulse.  The fluence range can also be seen in
  histograms of the burst fluence distribution, shown in Figure
  \ref{fig:energy hist}, again grouped by frequency. The apparent
  decay of the histograms  below $\sim 10^{-2}$Jy-s (1.2-1.7 GHz),
    or below $\sim 10^{-3}$Jy-s (4.5-5.0 GHz) is not physical, but is
  due to incomplete sampling of weak pulses.  The fluence at which the
  histograms peak corresponds approximately to the threshold we used
  to record a single pulse (discussed in \S \ref{sec:obs}).  The
  high-fluence side of the histograms is physical, however, reflecting
  the true distribution of burst fluence above our trigger threshold.

  The steepening of the burst fluence distribution at high fluence
  values is also consistent with previous statistical studies of
  single pulses from the Crab pulsar.  Several authors (Moffett 1997,
  Popov \& Stappers 2007, Bhat et al.\ 2008, Sheckard et al.\ 2010,
   Karuppusamy et al.\ 2010) used observations at lower time
  resolution to characterize the fluence distribution of total pulses
  (not separated into microbursts).  While details of the methods
  differ, all seem to agree that the total pulse fluence distribution at 1.4
  GHz steepens around $\sim 3\!-\!10 \times 10^{-3}$ Jy-s.  Thus, at
  the higher fluence values which our trigger threshold allowed us to
  sample, there is evidence that the fluence distribution is not a
  simple power law, but is in fact a convex function, steepening
  towards higher fluence.  Although we are measuring burst fluence,
  not total pulse fluence, most pulses contain no more than one very
  strong burst.  The high-fluence shape of our burst distribution
  therefore appears to be consistent with these previous studies.

   Figure \ref{fig:flux-width} also shows a tendency for brighter
  bursts to be shorter-lived, and for  weaker bursts to be 
  longer-lived.  This is consistent with previous results on single
  pulses (not bursts) seen at lower time resolution.  Bhat et al.\ (2008)
  compared the strength of a pulse to its effective width, defined
  in a matched-filtering sense;   Karuppusamy et al.\ (2010) compared
  the strength of a pulse to its equivalent width.
  Both papers found that brighter pulses
  tend to be shorter-lived.

To connect our observations to the range of energies released in a
burst, we must convert the observed fluence to energy.  As discussed
in \S\ref{sec:fits}, the energy of a burst is related to its fluence
by $E = {\cal E} \Delta \nu D^2 \theta^2$.  We know the distance to
the pulsar ($D = 2$ kpc); from our simultaneous two-frequency
observations (\S \ref{sec:two freq}) we can estimate $\Delta \nu \sim
0.5 \nu$.  Our typical burst fluence at 1.4 GHz is ${\cal E} \sim
10^{-2}$ Jy-s, and a factor $\sim 10$ lower at 5 GHz, but burst
fluences can vary by a factor $\sim 10$ up or down from these values.
To convert to energy, we guess that a burst is beamed into an angle
$\theta \sim 1 / \gamma$, where $\gamma$ is the Lorentz factor of the
emitting plasma.  This puts the energy released at the star at $\sim
(10^{26}\!-\!10^{28})/\gamma^2$ erg in the bursts we detected at 1.4 GHz.
In addition, it is very likely that many weaker bursts exist below our
detection threshold.

\subsection{What can we learn from microburst strengths?}
\label{subsec:energies}

We are not aware of many models which can be compared to our observed
burst energies; the nonlinear physics involved in coherent radio
emission make such predictions challenging.  One attractive idea which
can be tested is the suggestion that bursts represent an energy
storage and release mechanism, so that all bursts would release the
same energy at the star (e.g., Benford 2003).  The distribution
of points in Figure \ref{fig:flux-width} is suggestive of an
energy-conserving relationship;  stronger bursts tend to be shorter-lived,
and weaker bursts tend to be longer-lived.  However, the situation is
not so simple, because observed bursts occupy a range $\sim 100$ in
fluence, which seems to suggest a wide range in burst energy.  Nonetheless,
it may be that the bursts all have the same energy in the rest frame of
the emitting plasma, and that their observed fluence range is due to
relativistic beaming.  We explore two simple versions of this idea.

One possibility is that the Lorentz factor of the emitting plasma is
constant for all bursts, but the angle between the plasma motion and
the line of sight, $\theta$, varies burst to burst. This is easy to
test. If $E_0$ is the burst energy measured in the plasma rest frame,
we observe energy $E = E_0 / \gamma ( 1 - \beta \mu)$, for $\mu = \cos
\theta$.  If the bursts are emitted isotropically in the plasma rest
frame, we will observe a burst energy distribution, $n(E) \propto d \mu /
dE \propto 1 / E^2$. Because ${\cal E} \propto E$ in this scenario, we
should see a fluence distribution, $n({\cal E}) \propto 1 / {\cal
  E}^2$.  We illustrate this prediction with a dashed line in the
fluence histograms of Figure \ref{fig:energy hist}.  For well-sampled
frequencies (1.2 to 5\,GHz), the figures show that a small range of
the high-${\cal E}$ distribution does appear consistent with the
$1/{\cal E}^2$ prediction. However, for stronger bursts the
distribution drops more rapidly than $1/{\cal E}^2$.  We thus conclude
that this hypothesis fails: microbursts are not all emitted with the
same energy from plasma moving at a single Lorentz factor.

Alternatively, the observed fluence range may be caused by the Lorentz
factor, $\gamma$, varying from burst to burst.  For angles within the
beaming cone, $\theta \ltw 1 / \gamma$, the factor $1 - \beta \mu
\simeq 1 / 2 \gamma^2$.  The observed burst energy becomes $E \sim 2
\gamma E_0$, and the observed fluence becomes ${\cal E} \sim 2
\gamma^3 E_0$.  (This strong dependence on $\gamma$ reflects the
change in the beaming angle as well as the simple Doppler boost of the
burst energy).  Thus, a small variation in $\gamma$, by a factor $\ltw
5$, can lead to the observed factor $\sim 100$ variation in fluence
that we observe.  If such fluctuations do exist in the radio-loud
region, then the energy-storage-and-release picture may be
correct, and the  $n({\cal E})$ distribution  reflects the underlying
(and unknown) $n(\gamma)$ distribution.  Because current models of the
radio-loud region generally invoke steady flow in the emitting plasma,
we cannot identify a simple model prediction to test against our data.
However, given the dramatic fluctuations we see in the strength of the
radio emission, we see no reason to assume steady flow; turbulent,
unsteady flow with the necessary amplitude may well exist in the
radio-loud region.

\subsection{Strength of bursts seen at two frequencies}

Figures \ref{fig:flux-width} and \ref{fig:energy hist} show that
microbursts tend to be weaker at higher frequencies.  This 
 can be seen from
the upper envelope of the $(F,W)$ plots, which
decreases from $\sim 0.3$ Jy-s at 0.33 GHz to $\sim 0.03$ Jy-s at 4.8
GHz.  This trend is generally consistent with the steep spectrum of the Crab
pulsar's mean profile  (e.g.,  Lorimer et al.\ 1995),  
 but hard to specify due to incomplete sampling at lower  fluence.

\begin{figure}[]
{\center
\includegraphics[width=0.8\textwidth,angle=90]{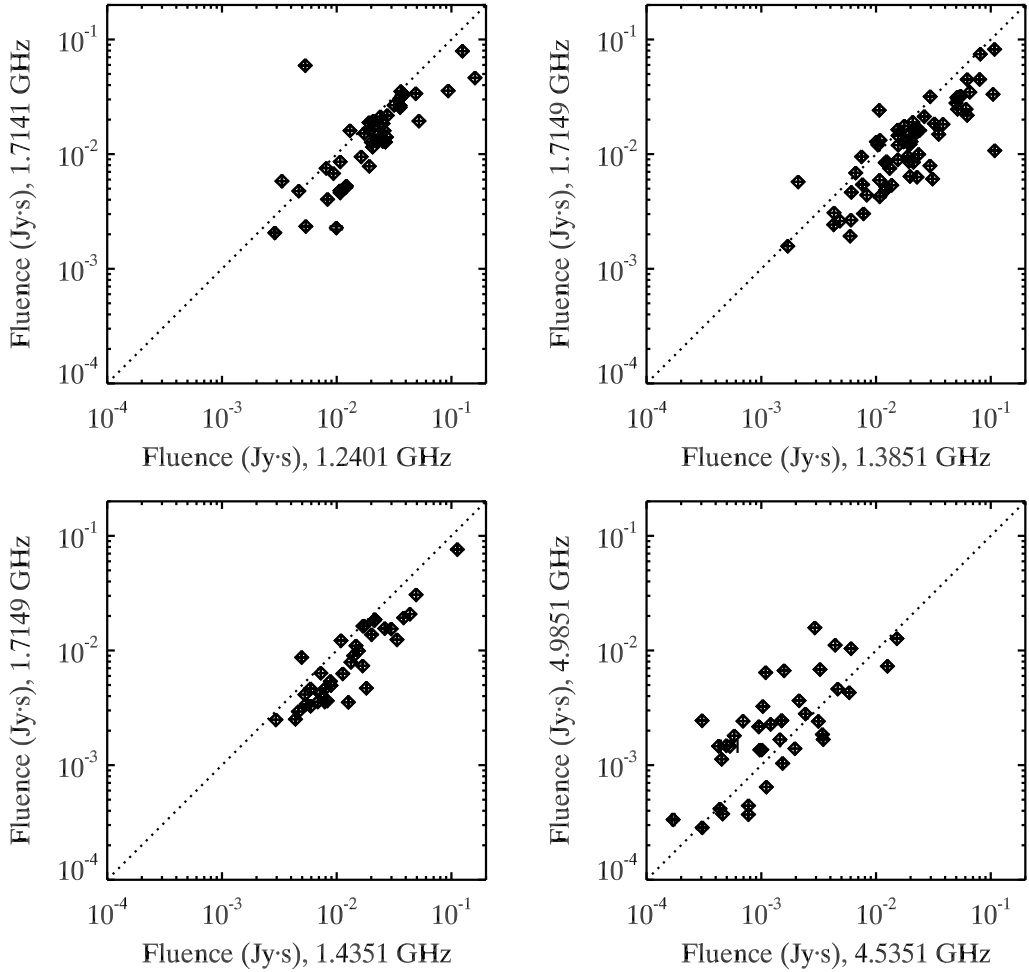}
\caption{Fluences of individual microbursts measured simultaneously at
two frequencies.  The dotted lines
  show the location of high- and low-frequency fluence equality.
  Error bars determined during the burst fitting process are plotted
  for both axes;  in all cases the error bars are as small or smaller than
the plot symbol.   The large scatter is physical, reflecting variation
of burst-to-burst fluence.  It is, nonetheless, apparent that
bursts measured between 1 and 2 GHz tend to be stronger 
at the lower measured frequency. }
\label{fig:two frequency:energy} }
\end{figure}

We can, however, be more quantitative if we restrict ourselves to
individual microbursts identified at both frequencies in pulses
simultaneously sampled at two frequencies.   From these data sets
we extract a total of 192 microbursts, which we can use to explore 
how the  fluence of an individual microburst depends on frequency.
Figure \ref{fig:two frequency:energy} compares the fluence of each burst at the
two observed frequencies.  Clearly the burst  fluences are well
correlated at the two frequencies.  The bursts are usually weaker at
the higher frequency, and show an approximately linear relationship
between the  fluence at the higher and lower frequency.  The scatter
in the figure is real, and substantially larger than errors from our
fitting procedure.

We quantify the fluence-frequency correlation by measuring the spectral
index, $\alpha$, for each burst, defined as ${\cal E}(\nu) \propto
\nu^\alpha$.  Table \ref{tab:width/energy} shows our results,
expressed as the mean $\langle \alpha \rangle$ and its standard
deviation, $\sigma_{\alpha}$.  Thus, our data suggest $\langle \alpha
\rangle \sim -2 $ for these microbursts, but with significant
uncertainty.  The relatively large scatter in the ${\cal
  E}(\nu_2)/{\cal E}(\nu_1)$
ratio for the set of bursts translates to the large uncertainty in the
derived spectral indices (especially at the very small frequency
separation between 4.5 and 4.9 GHz).  It should be noted that our
observing method introduces a bias for stronger pulses at higher
frequency, because the fluence threshold used to 
record individual pulses is compared only with the
  higher frequency pulse, which arrives first at our detector.  Our
results are consistent with those of Moffett (1997),
who measured a distribution of single pulse spectral indices between
1.4 and 4.9 GHz which peaked in the range $-2.0 < \alpha < -1.5$.
% $ \alpha \sim (-2.0, -1.5)$.

\section{Microburst duration}
\label{sec:widths}

Microbursts vary significantly in duration (width), even within a
single observing epoch at a single frequency.  Some are shorter-lived
(narrow), others are longer-lived (broad).  Unlike the situation with
burst energies, models do exist for the temporal structure of pulsed
emission.  At low frequencies, the observed pulse width is commonly
thought to be caused by interstellar scattering (ISS).  At higher
frequencies, where ISS effects are small, microstructure models (as in
\S \ref{sec:intro:micro}) attempt to connect temporal variability to
conditions in the pulsar's magnetosphere.  Our data allow us to verify
the former idea, learn at what frequencies intrinsic broadening
dominates ISS for the Crab pulsar, and establish properties of burst
widths which future models must address.

\subsection{Low frequencies:  pulse broadening by ISS}
\label{subsec:intro_to_widths}

 At low frequencies, pulse widths --- and by extension microburst
  widths --- have generally 
been ascribed to ISS caused by turbulence in the Crab Nebula or the
interstellar medium (ISM).  In particular, several authors have
studied pulse profiles of the Crab pulsar at frequencies below 1 GHz,
either as mean profiles or single pulses.  These authors find that
pulses at these frequencies are broad, single bursts, with a
consistent width and shape pulse to pulse, as would be expected from
ISS.  Our results at 0.33 GHz agree with this trend.  We find only one
burst per pulse, with little variation of burst width about the 
$\sim$ 600 $\mu$s mean.

\begin{figure}[]
{\center
\includegraphics[width=0.8\textwidth]{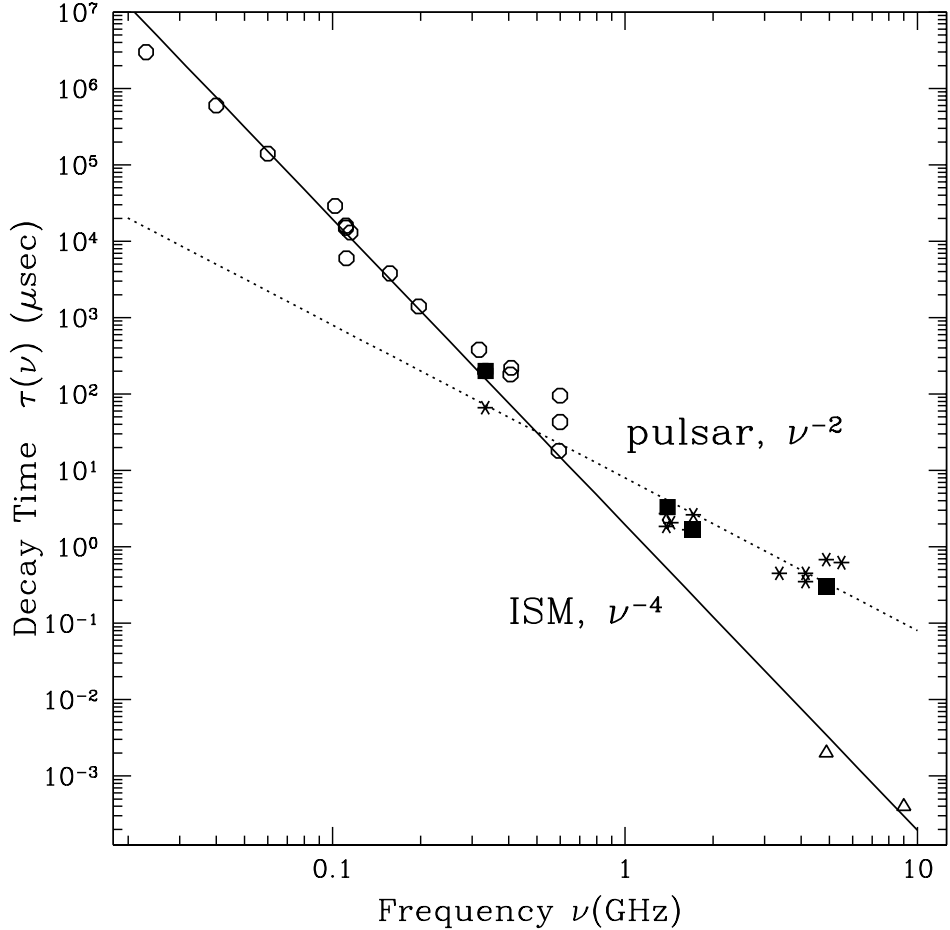}
\caption{Pulse decay time as a function of frequency.   Open circles show
  decay times  from the literature, listed in Table \ref{tab:scatt
    times}. Filled squares are estimates of the ``typical''
  microburst widths from our data, grouped by frequency as in Figure 
\ref{fig:flux-width}, and  converted to decay times using $\tau = W/3$,
 as discussed in \S \ref{sec:width-stats}.  
 Asterisks are decay times derived from mean fluctuation
  spectra, as listed in Table \ref{tab:fluc spectra}.  Triangles show upper
  limits on nanoshot widths from Hankins et al.\ (2003), at 4.8 GHz, and
Hankins \& Eilek (2007), at 9 GHz.  The solid line illustrates the 
  $\tau(\nu) \propto \nu^{ -4}$ behavior  predicted by interstellar
  scattering on Gaussian turbulence.  This line is not a formal fit to 
the points,   but simply meant to illustrate the trend.  The lower-frequency
points from the literature (open circles), and the highest-frequency 
nanoshot points (triangles) appear consistent with the ISS line. 
However, the characteristic widths
of our microbursts above 1 GHz (filled squares and asterisks) deviate
dramatically from the ISS prediction.  They are approximately consistent with
a $\tau(\nu) \propto \nu^{-2}$ law, which we illustrate with a dotted
line.}
\label{fig:the width plot} }
\end{figure}

In Table \ref{tab:scatt times} we combine our results at 0.33 GHz with
results from the literature; we show these results as open circles in
Figure \ref{fig:the width plot}.  Results in the literature are
typically quoted in terms of the exponential decay time $\tau$
(related to our widths by $\tau = W/3$).  We also overlay a $\tau(\nu)
\propto \nu^{-4}$ line in Figure \ref{fig:the width plot}, to
illustrate the width behavior predicted if the turbulence causing the
ISS has a Gaussian spectrum.  We note that small variations in the exponent
of the $\tau(\nu)$ behavior have also been suggested.  If the
turbulence has a Kolmogorov spectrum, models predict $\tau(\nu)
\propto \nu^{-4.4}$ (Lee \& Jokipii, 1975).  Kuzmin et al.\ (2002) fit
$\tau(\nu) \propto \nu^{-3.8}$ to data between $0.04\!-\!2.23$
GHz. 

Figure \ref{fig:the width plot} shows that the data below $\sim 1$ GHz
 are generally consistent with the
$\nu^{-4}$ prediction (or possibly with one of its variants).
The pulse widths below 1 GHz do show some scatter about the predicted
$\nu^{-4}$ line.  Some of
this may be introduced by the fact that these measurements, taken from
the literature, are not contemporaneous.  The mean-profile pulse width
of the Crab pulsar is known to vary on timescales of weeks or months;
this is believed to be due to ``weather'' in the Crab Nebula (e.g.,
Isaacman \& Rankin 1977).  Extreme examples are the strong scattering
events reported by Lyne \& Thorne (1975), or Backer, Wong \& Valanju
(2000).  To our knowledge the data in Table \ref{tab:scatt times} were
not obtained during such extreme events, but some secular variation
may still be expected.

\subsection{High frequencies:  burst widths are intrinsic}

Taken altogether, the data discussed in \S \ref{subsec:intro_to_widths}
support the idea that pulses seen below $\sim 1$
GHz are broadened by turbulence in the Crab Nebula and the ISM.  At higher
frequencies, however, the situation is very different.

Just as the energies of microbursts seen above $\sim 1$ GHz
 can vary over more than an order of
magnitude within one pulse, on time scales $\sim 10\!-\!100\mu$s, so
can their  durations. This can be seen by inspection of the
example pulses above $\sim 1$ GHz
 in Figure \ref{fig:fit example}, which show that bursts
within a single pulse can have vastly different widths.
Both narrow and broad bursts can coexist and
overlap in the same pulse. The burst widths we measured
range over nearly two orders of magnitude. 
This is apparent in Figure \ref{fig:flux-width}, and  also in Figure
\ref{fig:width hist}, which shows histograms of the burst widths, $n(W)$.

Microbursts seen above $\sim 1$ GHz  are almost always
longer-lived than the duration predicted by ISS.  Taking 600 $\mu$s as
a typical width at 0.33 GHz, and extrapolating as $W(\nu) \propto
\nu^{-4}$, we would expect widths determined by ISS to be $\sim$ 1
$\mu$s at 1.4 GHz.  For comparison, 
the distribution of our measured widths  peaks at $\sim$ 10 $\mu$s.  
At 4.8 GHz, the distribution of our measured widths peaks
at $\sim$ 1 $\mu$s; but we would expect widths due to  
ISS to be only $\sim 10$ ns. 
Once in awhile, however, the Crab pulsar  emits very narrow nanoshots
which are sparse enough in time to be detected individually.  Hankins
et al.\ (2003) detected a few nanoshots at 5 GHz which were unresolved
with 2 ns instrumental resolution.  Hankins \& Eilek (2007) detected a
nanoshot at 9 GHz which was unresolved at 0.4 ns resolution.  We
include these two results as open triangles in Figure \ref{fig:the
  width plot}, and note they fall approximately on the $\nu^{-4}$ ISS
line.  The duration of these nanoshots is consistent with the
high-frequency extrapolation of ISS.   Thus, individual nanoshots are
sufficiently  short-lived that they are subject to pulse broadening by ISS.
  However, the duration of the ``clumps'' of overlapping nanoshots
  which comprise most microbursts is too long to be due to ISS.

\subsection{High frequencies:  is nebular scattering important?}
\label{subsec:origin}

 Because bursts seen above $\sim 1$ GHz last longer than predicted
  by ISS, and because their widths can vary in less than a
  millisecond, we suspect their durations are intrinsic to the pulsar.
  Other authors (Cordes \& Lazio 2001, Karuppusamy et al.\ 2010) have
  disagreed, suggesting that scattering by structures in the Crab
  Nebula can account for pulse or burst widths even at high
  frequencies.  We have, in fact, seen two likely examples of nebular
  scattering.  On two observing days we saw strong, narrow bursts
  which were consistently followed by broad, weak bursts.  We
  identified the weak bursts as ``echoes'' of the strong ``primary''
  bursts (Crossley et al.\ 2007).  In each case the echoes persisted
  for $10^4\!-\!10^5$ stellar rotations, at a steady $\sim 50\!-\!
  100 \mu$s time lag and steady $\sim 1/3$ fluence ratio relative to
  the primary bursts.  We suggested in Crossley et al.\ (2007) that
  these echoes may be caused by structures in the Crab Nebula which
  happen to cross the sightline to the pulsar.

 Our two echo observations were unusual, however. Most of the
  microbursts we  recorded seem to occur randomly in duration,
  rotation phase and amplitude.  In particular, burst widths can vary
  dramatically within one stellar rotation, and sometimes within a
  single pulse.  To explain this by nebular scattering,
different scattering   clouds  must 
 cross our sightline every few milliseconds.  Such clouds
  must be very small, no larger than $L_{cl} \sim 300$ km.  Turbulent
  broadening from clouds this small  almost certainly occurs in the
  confined-screen limit of Cordes \& Lazio (2001).  In this limit,
the duration of the
  scattered burst is limited by the spatial extent of the cloud, not
  the turbulence level within the cloud.  For a cloud at distance
  $D_{cl}$ from the pulsar, the scattering width $\tau_{sc} \sim
  L_{cl}^2/ 2 c D_{cl}$.  To compare the predicted $\tau_{sc}$ to our
  observed widths, we must know $D_{cl}$. If the scattering cloud is associated
  with the Crab nebula, rather than the pulsar, it must be 
at or beyond the termination shock of the pulsar wind.  
Momentum balance says this shock 
 should occur at $\sim 0.1$ pc, which 
  coincides nicely with the location of the quasi-stationary X-ray
  ``ring'' believed to be associated with that shock (e.g.,
  Hester 2008).  But now, the scattering width caused by a cloud 300 km in
  size, sitting at $ D_{cl} \sim 0.1$ pc, is only $t_{sc} \sim
  10^{-14}$s.  Clearly this simple model cannot explain the burst durations
  we observe.

 Alternatively, to broaden a narrow burst to $\sim 10 \mu$s, a
  300-km cloud must be located at $D_{cl} \sim 10^4$ km from the
  pulsar.  This distance is about 10 times the light cylinder radius, but
well inside of the termination shock;  thus the cloud is 
within the pulsar wind.  Such small structures may exist in a pulsar
wind, but their properties (size, magnetization, plasma content)
  are not well enough understood to expore this idea any further.

 The nebular-scattering model is also called into question by the
  lack of clear evidence for turbulent scattering in burst profiles
  above $\sim 4$ GHz.  Turbulent scattering creates an exponential
  tail on the scattered burst; but the bursts we recorded at $4 \!-\!
  5$ GHz tend to have a symmetric profile (as discussed in \S
  \ref{sec:fit_robust}).  The same is true for microbursts Hankins \&
  Eilek (2007) detected in the main pulse between 6 and 10 GHz. It
  therefore seems likely that burst profiles above $\sim 4$ GHz
  reflect the temporal behavior of the fundamental energy-release
  process. On the other hand, bursts we recorded between $ 1 \! -
  \! 2 $ GHz do have a fast-rise, slow-decay shape.  Their profile may
  be determined solely by the energy-release process, if that process
  has a different temporal signature than those which create
  higher-frequency bursts.  The burst profile at
  $1\!-\!2$ GHz may also be modified by scattering; if that is the case,
our arguments above suggest that the scattering must happen no further
from the pulsar than within its wind. Overall, it seems simplest to
 argue that all burst durations
above $\sim 1$ GHz are intrinsic to the pulsar.

\subsection{Frequency dependence of burst widths}
\label{burst_width_freq}

Despite the wide range of burst durations we observed, 
figures \ref{fig:flux-width} and \ref{fig:width hist} show that the 
`` typical''  width of a burst depends on frequency.   Microbursts seen at
higher frequency tend to be shorter-lived than those seen at lower frequency.   
We can quantify this using bursts observed simultaneously at two
 frequencies, as well as with statistical measures of our full data sets.

\subsubsection{Individual bursts seen at two frequencies}

\begin{figure}[]
{\center
\includegraphics[width=0.8\textwidth,angle=90]{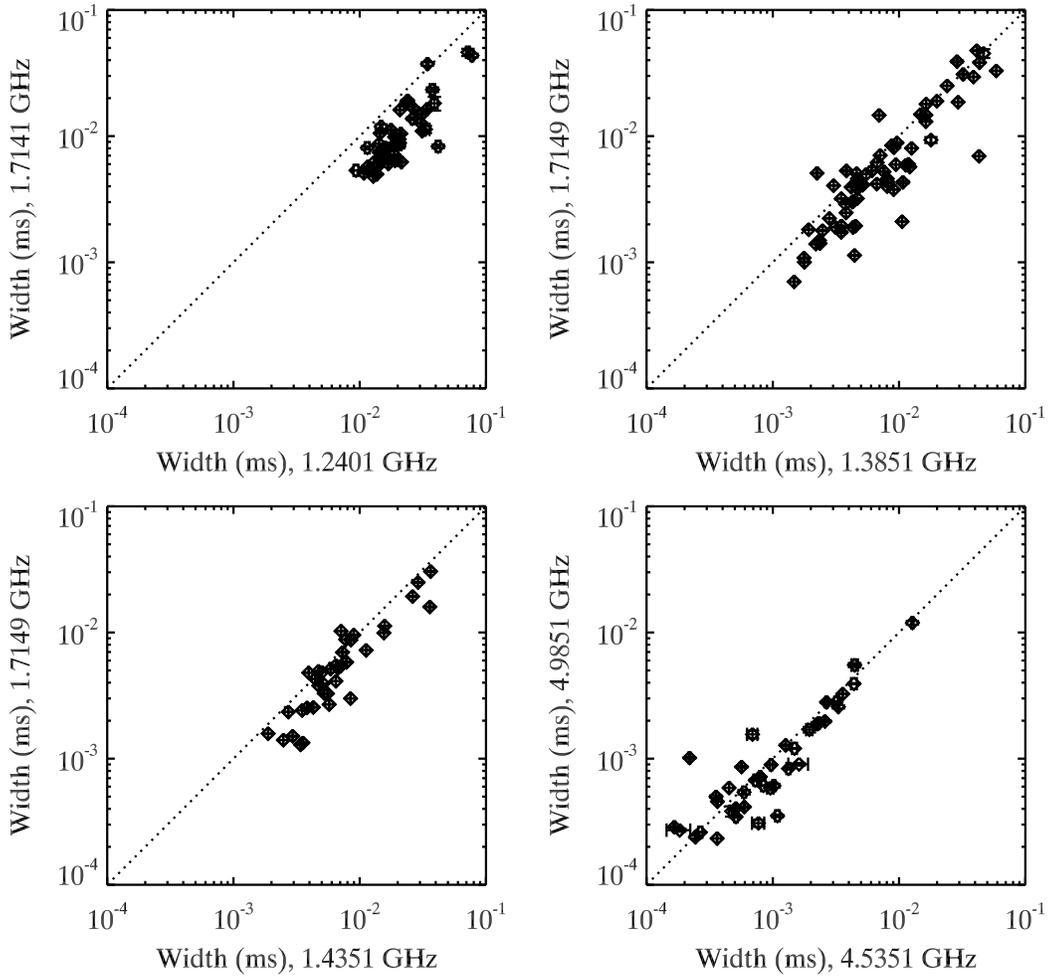}
\caption{Widths of individual
microbursts measured simultaneously at two frequencies.
The dotted line shows the location of high-  and low-frequency width equality.
  Error bars
  determined during the microburst fitting process are plotted for
  both axes; for most points the error bars are smaller than the plot
symbols.   The scatter is physical, reflecting burst-to-burst
variation in widths. It is apparent that microbursts seen between
1 and 2 GHz tend to be broader
when observed at the lower frequency.}
\label{fig:two frequency:width} }
\end{figure}

As with burst energies, we can use simultaneous two-frequency
observations to explore the frequency dependence of burst duration.
Figure \ref{fig:two frequency:width} compares the width of each burst at
the two observed frequencies.  We find that burst
widths at the two frequencies are well correlated, but not identical.
Lower frequency microbursts are typically longer-lived than their higher
frequency counterparts, but again we find strong scatter about the
correlation.  As with the energy-energy correlations, the scatter in
width-width correlations is real, and substantially larger than the
errors in our fitting procedure.  We quantify the width-frequency
correlation by assuming the width obeys $W(\nu) \propto \nu^\beta$,
where $\beta$ is the width index.  Table \ref{tab:width/energy} shows
our results, expressed as the mean $\langle \beta\rangle$ and its
standard deviation, $\sigma_{\beta}$.  Our data suggest $\langle \beta
\rangle \sim -2$, again with significant uncertainty (as shown in
the $\sigma_{\beta}$ values).

\subsubsection{Statistical measures of burst widths}  
\label{sec:width-stats}

\begin{figure}[]
{\center
\includegraphics[width=0.7\textwidth,angle=90]{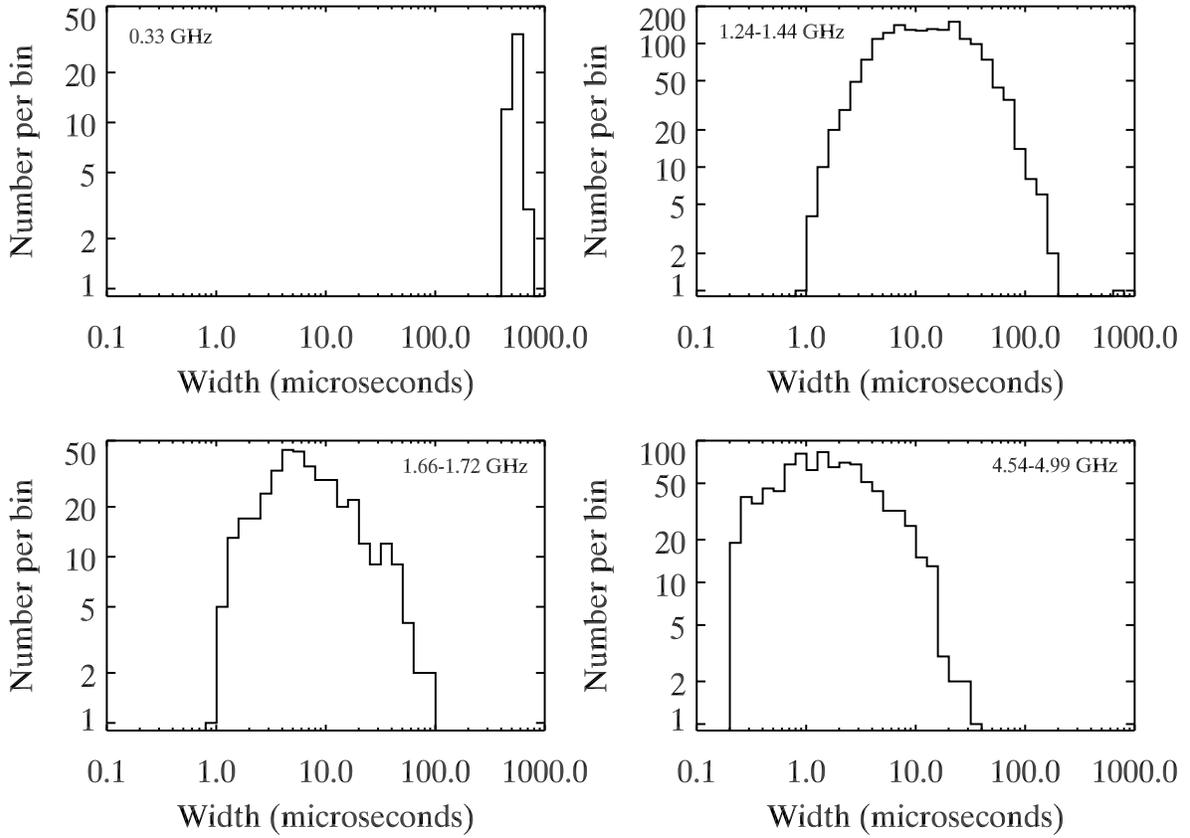}
\caption{Distribution of widths for all measured microbursts,
displayed as a histogram in log space, grouped in the same frequency ranges
used in Figures \ref{fig:flux-width} and \ref{fig:energy hist}. These
data are binned in logarithmic intervals, as in Figure \ref{fig:energy hist}. 
At 0.33 GHz, the pulse widths are dominated by ISS.  At higher frequencies,
the broad width distribution is intrinsic to the pulsar, as discussed in 
\S \ref{sec:widths}. }
\label{fig:width hist} }
\end{figure}

We can also use our full set of measured bursts to
 investigate the $W(\nu)$ relation.  We first simply estimate the mean
 width, $\langle W \rangle$, described by the the $n(W)$ distributions 
 in Figure \ref{fig:width hist}.  Because the width at which these
 histograms peak is not an artifact of our observation or analysis
 methods, but is intrinsic to the star, it is an accurate
 characterization of microburst durations.  At 0.33 GHz, we estimate
 $\langle W \rangle \sim$ 600 $\mu$s; $ \sim$
 10 $\mu$s at 1.4 GHz; $\sim$ 5 $\mu$s at 1.7 GHz; and $\langle W
 \rangle \sim$ 1 $\mu$s at 4.8 GHz.  These results are included
 as filled squares in Figure \ref{fig:the width plot}, where we again convert
to decay times: $\langle \tau \rangle \sim$ 200 $\mu$s at 0.33 GHz; $ \sim$
 3 $\mu$s at 1.4 GHz; $\sim$ 1.7 $\mu$s at 1.7 GHz; and $\langle \tau
 \rangle \sim$ 0.3 $\mu$s at 4.8 GHz.

 For an alternative analysis, we might want to determine a mean burst
 width from each of our data sets.  Because of the intrinsic jitter of
 the microbursts in pulse phase, we cannot simply average a set of
 pulses. That would give a mean pulse profile (or a form of the
 probability density function of microbursts in pulse phase) but not a
 mean burst profile.  Instead, we work in the Fourier domain, where
 delays in arrival time are only phase shifts.  When the modulus of
 the Fourier components is computed, the phase shifts are irrelevant.
 We therefore evaluate the mean fluctuation power spectrum of each set
 of pulses.

Our method is as follows. For each pulse in a data set, we use a Fast
Fourier Transform to evaluate the fluctuation power spectra for both
an on-pulse and an off-pulse region, letting $f$ be the conjugate
Fourier variable to time $t$.   In the off-pulse region, multiple
transforms are averaged to decrease the estimation error.  The
on-pulse spectrum is divided by the off-pulse spectrum to remove any
(non-white) residual spectrum from the receiver bandpass.  All of the
corrected on-pulse spectra for a given observing session are
normalized (to prevent a single strong pulse from dominating the
result) and averaged (to derive a mean fluctuation spectrum).

In order to determine the characteristic timescale of the bursts,
we  fit an analytic function to the discretized mean fluctuation spectrum. 
We chose a generalized burst shape in the time domain, 
\begin{equation}
G_n(t) = t^n e^{ - t / \tau}
\end{equation}
If $n + 1 > 0$, $G_n(t)$ has an analytic Fourier transform, $g_n(f)$;  its
squared amplitude is
\begin{equation}
\left| g_n(f) \right|^2 = 
{ \Gamma^2(n+1) \over \left( 4 \pi^2 f^2 + 1 / \tau^2 \right)^{n+1}}
\label{FlucModel}
\end{equation}
where $\Gamma^2(n+1) = (n!)^2$ is the square of the usual Gamma function.
We compared equation (\ref{FlucModel}) to the mean 
fluctation spectrum of  11 data sets, recorded at the VLA (picked from the
Table \ref{tab:VLA obs table}), and at Arecibo (from Table \ref{tab:AO obs
  table}), chosen to sample our frequency range well.  We kept $n$ as a fixed 
parameter, and used least-squares methods to determine the value of $\tau$ for
each data set.

\begin{figure}[]
  {\center 
\includegraphics[width=0.95\textwidth]{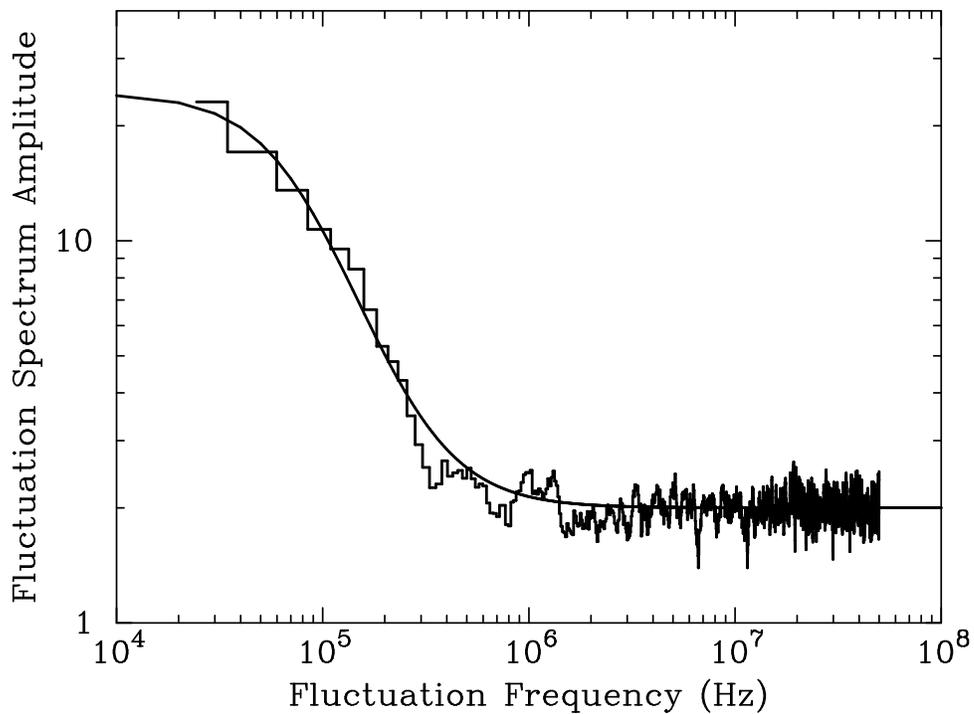} 
    \caption{The mean, on-pulse fluctuation spectrum for a set of 34 pulses at 
1.4351 GHz  (MJD 50368).  The solid line shows the best-fit estimate 
of the $n=0$ template, $g_{0}(f)$, for which $\tau = 2.1~ \mu$s.  Because the fluctuation
spectrum of each pulse was normalized before forming the mean, the vertical scale is
arbitrary.}
\label{fig:fluct_spec_Lband} }
\end{figure}

\begin{figure}[]
{\center
\includegraphics[width=0.95\textwidth]{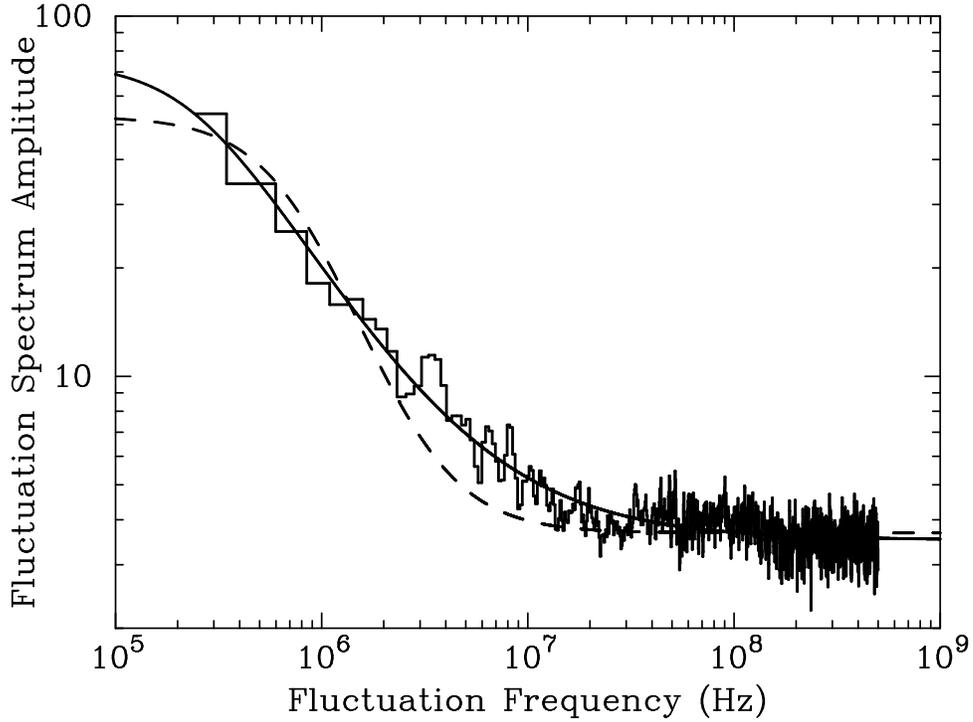}
\caption{The on-pulse fluctuation spectrum for a set of 15
  pulses at 5.5 GHz (MJD 52336). The solid line shows the best-fit estimate
of the $n=-1/2$   template, $g_{-1/2}(f)$, for which $\tau = 0.62~\mu$s.  The
dotted line shows the best-fit $n = 0$ template, which clearly does not
fit these data well. Other details are as in Figure 
  \ref{fig:fluct_spec_Lband}; when compared with that figure the difference in
  the structure of the fluctuation spectrum is clear.  }
\label{fig:fluct_spec_Cband} }
\end{figure}

Although the $g_1(f)$ case recovers the function we used to fit
individual bursts (equation \ref{xfred}), we found that this case does
not fit any of our mean fluctuation spectra well.  This is perhaps
surprising, because our fitting function is well matched to the shape
of individual microbursts, especially below 2 GHz.  Nonetheless, we
found that mean fluctuation spectra below 1.7 GHz are well fit by the
simpler $g_0(f)$ function, which is the transform of a one-sided
exponential decay.  Figure \ref{fig:fluct_spec_Lband} shows an example
of our fit to a fluctuation spectrum at 1.4 GHz.  However, fluctuation
spectra above 1.7 GHz were not well fit by $g_0(f)$ or $g_1(f)$.
After experimentation we found that $g_{-1/2}(f)$ provides a better
description of the mean fluctuation spectra at the higher
frequencies. Figure \ref{fig:fluct_spec_Cband} shows an example of
this fit at 5.5 GHz, and also illustrates how poorly $g_{0}(f)$
matches these data.  Because the corresponding time-domain function,
$G_{-1/2}(t)$, diverges as $t \to 0$, we do not claim it is a true
description of a ``characteristic'' microburst at all times.  We
suspect this apparently unphysical result comes from combining
normalized fluctuation spectra of broad and narrow microbursts into
one mean spectrum.  We simply note that $g_{-1/2}(f)$ is a good
representation of the data for our mean fluctuation spectra above 2
GHz, and use this representation to estimate the mean decay constants
for those data sets.

Our results are summarized in Table \ref{tab:fluc spectra}, and
included as asterisks in Figure \ref{fig:the width plot}.  We were
pleased to find that these mean decay constants are comparable to the
simple estimates of burst duration which we made from the $n(W)$
histograms in Figure \ref{fig:width hist}, even though they were
derived by quite different methods.  Both of these methods also agree
with the width behavior we inferred from individual
 bursts seen at two frequencies.  We 
conclude that the typical widths of microbursts seen above $\sim 1$ GHz obey
 $W(\nu) \propto \nu^{-2}$, but that 
 significant burst-to-burst scatter also exists about this trend.

\subsection{What can we learn from microburst widths?}
\label{subsec:widths}

We argued in the previous section that the duration of each microburst
seen above $\sim 1$ GHz is set when it leaves the pulsar. 
The burst width must be caused either by the emission process, or by propagation
through the star's magnetosphere, or both.  If this is the case, then
our results provide simple constraints on different microstructure
models, as follows.

One model holds that microbursts are due to long-lived geometrical
structures or narrow radiation beams within the magnetosphere,
(as discussed in \S \ref{sec:intro:micro}).  
If the bursts are due to narrow flux tubes,
radiating within an angle $\sim 1 / \gamma$ as they rotate through the
line of sight, their duration is $\sim P / 2 \pi \gamma$, where $P$ is
the pulsar's rotation period.  Thus, a
range  $\sim 100$ in observed burst duration requires that a range 
$\sim 100$ in the plasma Lorentz factor coexists between different flux tubes
within the open field line region.  Such large
variation is not predicted by current models of plasma in the open
field line region.  Alternatively, if the 
bursts are due to small charge clouds moving out along a magnetic
field line with curvature radius $\rho$, their duration is $\sim \rho
/ \gamma^3 c$.  The observed range  $\sim 100$ in burst duration
requires a range $\ltw 5$ in the Lorentz factor of different clouds
observed within a single pulse.  We argued above, in \S\ref{subsec:energies},
 that velocity fluctuations of this magnitude may not be unreasonable
 in a turbulent, radio-loud region. 

Another type of model  attributes
microstructure to stimulated Compton scattering, and the sweep of a
narrow, exponentially-enhanced radiation beam past the line of sight
(e.g., Petrova 2004).  These models are attractive, in that they
 have the  potential to create widely varying pulse widths and energies,
 due to the extreme sensitivity of the exponential gain factor to local
plasma parameters. Unfortunately, it is hard to derive simple,
testable predictions from these models, due both to their complexity,
and to the intrinsic nonlinear behavior of the underlying physics.

A third type of model argues that microstructure reflects intrinsic
temporal variability of the underlying plasma.  One example here is
the hypothesis that bursts each release a constant energy, $E_0$,
as measured in the plasma rest frame (\S\ref{subsec:energies}).
Doppler beaming causes the fluence of such a burst to be ${\cal E}
\propto \gamma^3 E_0$.  If the bursts also have a constant rest-frame
duration,
 $\tau_0$, relativistic and light travel effects shorten their duration as seen at
earth to $\tau \sim \tau_0 / \gamma$.  Thus, 
the flux of the burst should obey $F \propto {\cal E}/\tau
\propto  \tau^4$.  We see no such correlation in Figure
\ref{fig:flux-width}.  We conclude that, while constant-energy bursts
with variable Lorentz factors can explain the range of fluences we
see, the intrinsic duration of such bursts cannot also be constant.

A variant of this model suggests that the burst energy, $E_0$, is
released by a reconnection event, of scale $L_0$ and local magnetic field
$B$, in a region of
turbulent plasma.  The duration of the event, as seen in the plasma
rest frame, is $\sim L_0 / c$ (noting that the reconnection
flow speed is close to lightspeed in the relativistic plasma of the
magnetosphere; Lyutikov 2003).  Light-travel effects shorten its
 duration as seen at earth to $\tau
\sim L_0/\gamma^2 c$.  The few-$\mu$s pulse widths we see at 1.4 GHz
therefore come from a region $L_0 \sim \gamma^2$ km.  Because such a
region must be smaller than the size of the
magnetosphere, this picture can work only if the radio-loud plasma is
moving fairly slowly ($\gamma \ltw 10$, say).  We expect
the energy released in a reconnection event to be $E_0 \propto B^2
L_0^3$, leading to a fluence ${\cal E} \propto \gamma^2 B^2 L_0^3$ as
seen at earth.  A spread in $L_0$ can cause the observed spread in
${\cal E}$.  If the pre-reconnection field $B$ is roughly constant
for each burst, the flux should obey $F \propto {\cal E}/\tau \propto
\tau^2$.  Once again, we see no such correlation in Figure
\ref{fig:flux-width}.  We conclude that this model can explain the
data only if the pre-reconnection magnetic field tends to be weaker in
larger (longer-lived) events.

\section{Summary and Discussion}
\label{sec:conclusions}

Our high-time-resolution observations reveal a wealth of structure within
single pulses from the main pulse of the Crab pulsar, against which
current and future models of pulsar radio emission and microstructure
should be tested.

At 0.3 GHz, the pulse profile is smeared and broadened by ISS, which
preserves the total pulse energy but broadens and distorts the profile
into a fast-rise, slow-decay shape.  We therefore  cannot determine
the intrinsic structure of a low-frequency pulse when it left the star.
 Above $\sim 1$  GHz, however, ISS effects become negligible, and
 the intrinsic pulse structure is revealed.  We find that individual
 pulses contain one to several microbursts, each burst lasting on the 
 order of microseconds. The bursts are often sparse enough
and bright enough to be measured
 individually.  We measured the duration, peak flux, and fluence
of  nearly 3000 bursts, at frequencies between 1.2 and 4.8 GHz,
recorded in  different observing sessions over several years.

Microbursts have only a modest bandwidth. When we carried out
simultaneous, two-frequency observations, we found that individual
bursts within a single pulse can be identified between 1.2 and 1.7
GHz, but not between 1.4 and 4.9 GHz.  Thus, while a particular
bright pulse can be broadband (detected over a factor $\gtw 2\!-\!3$ 
in frequency), the bursts it contains are relatively narrowband
($\Delta \nu / \langle \nu \rangle < 1$). 

Microbursts at a given frequency can be bright or faint.  There is
some tendency for high-flux bursts to be shorter-lived and low-flux
bursts to be longer-lived.  Although this trend could suggest that all
bursts release the same amount of energy, our measured
burst fluences varied over more than a factor $\sim 100$, and even
weaker bursts are very likely common but below our detection
threshold.  It may be that this spread is caused by relativistic
beaming of constant-energy bursts from a turbulent plasma in the radio-loud
region.  Unfortunately, 
models of the radio emission region are not yet developed to the
point where specific tests against the data are possible.

Microbursts at a given frequency can be short or long, but they
also tend to be briefer at higher frequency. 
Simultaneous two-frequency observations as
well as statistical estimates from our full data sets suggest that the
burst width becomes shorter with frequency approximately as $W(\nu)
\sim \nu^{-2}$. Because bursts above $\sim 4$ GHz do not show the
exponential tail characteristic of turbulent scattering,  and because
burst widths above $\sim 1$ GHz fluctuate on very short timescales,
we argue that the
burst duration is intrinsic to the energy release mechanism.
The large range of burst durations may be caused, at 
least in part, by relativistic boosting from turbulent plasma in the
radio emission region.  However, current models of microstructure are
not developed to the point where they can be quantitatively tested
against the data.

\begin{acknowledgements}

  We thank the technical, operations, and computer staffs at the Very
  Large Array and the Arecibo Observatory for their help with the data
  acquisition equipment, observing, and for providing the computing
  environment we used for the observations.  We also thank David
  Moffett, Tracey Delaney and Joe Dickerson for help with the VLA
  observations.   We are grateful to Jim Sheckard for many 
enlightening discussions of microstructure
  models,  to Jim Weatherall for suggesting the relativistic beaming
  angle modification of a constant-energy burst, and to Jim Cordes for
conversations about possible causes of 
pulse broadening.  We thank the referee for perceptive 
suggestions which improved the paper.  This work was partly
  supported by NSF grants AST-9618408, AST-0139641, and AST-0607492.

\end{acknowledgements}

\clearpage

% ----------------

% ----------------

% ----------------

% ----------------

%^^^^^^^^^^^^^^^ Table of VLA Observing Sessions ^^^^^^^^^^^^^^^^
\begin{deluxetable}{lcccccc}
%\tablecolumns{7}
\tablewidth{0pt}
\tablecaption{VLA observations used for microburst analysis}
\tablehead{
\colhead{Center }               & \colhead{Band-} & \colhead{No.}   & \colhead{No.}   & \colhead{Smooth-}  
    & \colhead{Time}   & \colhead{MJD} \\
\colhead{Freq\tablenotemark{1}} & \colhead{width} & \colhead{of}    & \colhead{of}    & \colhead{ing}      
       & \colhead{Span}   &               \\
\colhead{(GHz)}                 & \colhead{(GHz)} & \colhead{pulses}& \colhead{bursts}& \colhead{time }    
             & \colhead{(min)}  &               \\
                                &                 &                 &                 & \colhead{($\mu$s)} 
           &                  &                 }
\startdata
0.3339     & 0.003 &  \phn49 &  \phn49 &    20.0 &  \phn18 & 49406   \\ 
1.2401$^a$ & 0.025 &  \phn38 &  \phn61 & \phn1.6 &  \phn55 & 51218   \\ 
1.3851$^b$ & 0.500 &  \phn27 &  \phn52 & \phn0.8 &  \phn41 & 51159   \\ 
1.3851$^c$ & 0.500 &  \phn52 &  \phn97 & \phn0.8 &  \phn20 & 51174   \\ 
1.4149     & 0.500 &  \phn61 &     142 & \phn0.8 &  \phn26 & 49399   \\ 
1.4351     & 0.500 &  \phn52 &  \phn95 & \phn0.8 &  \phn14 & 51137   \\ 
1.4351     & 0.500 &  \phn50 &     131 & \phn0.8 &  \phn11 & 51159   \\ 
1.4351     & 0.500 &  \phn54 &     123 & \phn0.8 &  \phn15 & 51167   \\ 
1.4351     & 0.500 &  \phn51 &     119 & \phn0.8 &  \phn10 & 51174   \\ 
1.4351$^d$ & 0.500 &  \phn34 &  \phn81 & \phn0.8 &  \phn13 & 51174   \\ 
1.4351     & 0.500 &  \phn51 &  \phn81 & \phn0.8 &  \phn22 & 51214   \\ 
1.4351     & 0.500 &     203 &     259 & \phn0.8 &  \phn59 & 50193   \\ 
1.4351     & 0.500 &     103 &     200 & \phn0.8 &  \phn84 & 50224   \\ 
1.4351$^e$ & 0.500 &  \phn34 &  \phn69 & \phn0.8 &  \phn30 & 50368   \\ 
1.4351$^f$ & 0.500 &  \phn61 &     108 & \phn0.8 &     170 & 50370   \\ 
1.6649$^b$ & 0.500 &  \phn70 &     145 & \phn0.8 &  \phn41 & 51159   \\ 
1.7141$^a$ & 0.025 &  \phn38 &  \phn63 & \phn1.6 &  \phn55 & 51218   \\ 
1.7149$^c$ & 0.500 &  \phn52 &     101 & \phn0.8 &  \phn20 & 51174   \\ 
1.7149$^d$ & 0.500 &  \phn34 &  \phn74 & \phn0.8 &  \phn13 & 51174   \\ 
4.5351$^g$ & 0.500 &  \phn46 &  \phn96 & \phn0.2 &  \phn41 & 51167   \\ 
4.8851     & 0.500 &  \phn51 &     104 & \phn0.2 &  \phn57 & 49080   \\ 
4.8851$^e$ & 0.500 &  \phn40 &  \phn79 & \phn0.2 &  \phn31 & 50368   \\ 
4.8851$^f$ & 0.500 &  \phn62 &     117 & \phn0.2 &     170 & 50370   \\ 
4.8851     & 0.500 &  \phn34 &  \phn54 & \phn0.2 &  \phn16 & 51112   \\ 
4.8851     & 0.500 &     152 &     312 & \phn0.2 &     126 & 51112   \\ 
4.9851$^g$ & 0.500 &  \phn52 &     145 & \phn0.2 &  \phn42 & 51167   \\ 
\enddata
\tablenotetext{1}{Superscript pairs denote simultaneous 2-frequency observing sessions.} 
\label{tab:VLA obs table}
\end{deluxetable}
% ^^^^^^^^^^^^^^^^^^^^^^^^^^^^^^^^^^^^

%^^^^^^^^^^^^^^^ Table of AO Observing Sessions ^^^^^^^^^^^^^^^^
\begin{deluxetable}{cccc}
\tablewidth{0pt}
\tablecaption{Arecibo observations used for fluctuation spectra}
\tablehead{
\colhead{Center Freq} & \colhead{Bandwidth}  & \colhead{number of}  & \colhead{MJD} \\
\colhead{(GHz)}                        & \colhead{(MHz)}      & \colhead{pulses}     & }
\startdata
% SORT BY FREQUENCY
3.375 & 250 & 16 & 52399 \\ 
4.150 & 500 & 14 & 52334 \\ 
4.150 & 500 & 12 & 52335 \\ 
5.500 & 500 & 15 & 52336
\enddata
\label{tab:AO obs table}
\end{deluxetable}
% ^^^^^^^^^^^^^^^^^^^^^^^^^^^^^^^^^^^^

% ^^^^^^^^^^^^^^^^ WIDTH AND ENERGY EXPONENTS ^^^^^^^^^^^^^^^^^^^^
% This is table 3, try revision
\begin{deluxetable}{ccccccc}
\tablewidth{0pt}
\tablecaption{Fluence and width behavior of microbursts }
\tablehead{
\colhead{Freq 1} & \colhead{Freq 2} & \colhead{No.\ of} & 
\colhead{$\langle \alpha \rangle$\tablenotemark{\it a}} & 
\colhead{$\sigma_\alpha$\tablenotemark{\it a}} & 
\colhead{$\langle \beta \rangle$\tablenotemark{\it b}} & 
\colhead{$\sigma_\beta$\tablenotemark{\it b}} \\
 \colhead{(GHz)}  & \colhead{(GHz)}  & \colhead{bursts}    &&&&}

\startdata
1.2401 & 1.7141 & 46 & $-1.2$  & 1.8 & $-2.2$ & 0.9 \\
1.3851 & 1.7149 & 68 & $-2.4$  & 2.4 & $-1.6$ & 2.1 \\
1.4351 & 1.7149 & 37 & $-2.7$  & 2.0 & $-1.8$ & 1.8 \\
4.5351 & 4.9851 & 41 & \phs4.8 & 8.2 & $-0.8$ & 4.7 %\\
\enddata
\tablenotetext{\it a}{Mean and standard deviation of fluence index, where 
   ${\cal E} (\nu) \propto \nu^\alpha$.}
\tablenotetext{\it b}{Mean and standard deviation of width index, where 
      $W(\nu) \propto \nu^\beta$.} \\
\label{tab:width/energy} 
\end{deluxetable}
% End of table 3
% ^^^^^^^^^^^^^^^^^^^^^^^^^^^^^

% **** SCATTER BROADENING WIDTHS FROM THE LITERATURE****
% This is  table 4:
\begin{deluxetable}{cr@{.}ll}
\tablewidth{0pt}
\tablecaption{Pulse scattering decay times below 1 GHz}
\tablehead{
\colhead{Frequency} & \multicolumn{2}{c}{Decay time}  & \colhead{Reference} \\
\colhead{(GHz)}     & \multicolumn{2}{c}{$\tau$ (ms)} & \colhead{ } }   %    \\
\startdata
0.023  & 3000&$^a$    & Popov et al.\ (2006) \\
0.040  & 600&$^a$    & Kuzmin et al.\ (2002) \\
0.060  & 140&$^a$    & Kuzmin et al.\ (2002) \\
0.102  &  29&$^a$    & Kuzmin et al.\ (1996) \\
0.111  &  16&$^a$    & Kuzmin et al.\ (2002) \\
0.111  &  15&$^a$    & Popov et al.\ (2006) \\
0.112  &   6&$^a$    & Rankin et al.\ (1970) \\
0.115  &  13&$^a$    & Staelin \& Sutton (1970)\\
0.157  &   3&8$^a$   & Staelin \& Sutton (1970)\\
0.197  &   1&4$^a$   & Rankin et al.\ (1970) \\
0.317  &   0&38$^a$  & Rankin et al.\ (1970) \\
0.333  &   0&20$^b$  & this paper \\
0.408  &   0&22$^a$  & Rankin et al.\ (1970) \\
0.406  &   0&18$^a$  & Kuzmin et al.\ (2002) \\
0.594  &   0&018$^a$ & Kuzmin et al.\ (2002) \\
0.600  &   0&095$^b$ & Sallmen et al.\ (1999) \\
0.600  &   0&043$^a$ & Popov et al.\ (2006) \\
\enddata
\tablenotetext{\it a}{Estimated from mean profile; Rankin's values converted as $\tau = W/\sqrt{2}$.}
\tablenotetext{\it b}{Estimated from mean of individual profiles}
\label{tab:scatt times}
\end{deluxetable}
%End table 4

% **** JEFF's FLUC SPECTRA RESULTS******
\begin{deluxetable}{cccc}
\tablewidth{0pt}
\tablecaption{Characteristic decay times from fluctuation spectra}
\tablehead{
\colhead{Frequency}   & \colhead{Decay time}            & \colhead{Observ-} &  \colhead{MJD} \\
\colhead{(GHz)}      & \colhead{$\tau_{\rm char}$ ($\mu$s)}  & \colhead{atory}   &  }
\startdata
0.3330		& $^a66.0$ & VLA & 49406 \\
1.3851		& $^a2.72$ & VLA & 51159 \\
1.3851		& $^a1.85$ & VLA & 51174 \\
1.4351		& $^a2.07$  & VLA & 50368 \\
1.6649		& $^a1.67$ & VLA & 51159 \\
1.7149		& $^b2.63$ & VLA & 51174 \\
3.3750		& $^b0.45$ & AO & 52399 \\
4.1500		& $^b0.35$ & AO & 52334 \\
4.1500		& $^b0.45$ & AO & 52335 \\
4.8851		& $^b0.68$  & VLA & 50368\\
5.5000		& $^b0.62$ & AO & 52336 \\
\enddata
\tablenotetext{\it a}{Decay time determined using $n =0$ case}
\tablenotetext{\it b}{Decay time  determined using $n = - 1/2 $ case}
\label{tab:fluc spectra}
\end{deluxetable}
% **** ****
% ^^^^^^^^^^^^^^^^^^^^^^^^^^^^^^^^^^^^

\end{document}